\documentclass[useAMS,usenatbib]{mn2e}
\usepackage{graphics, graphicx}
\newif\ifAMStwofonts

\def\lapp{\ifmmode\stackrel{<}{_{\sim}}\else$\stackrel{<}{_{\sim}}$\fi}
\def\gapp{\ifmmode\stackrel{>}{_{\sim}}\else$\stackrel{>}{_{\sim}}$\fi}

\newcommand{\degrees}[1]{\ensuremath{#1^\circ}}

\title[Interpulse pulsars and beam evolution]
{The population of pulsars with interpulses and the implications for beam evolution}
\author[Weltevrede \& Johnston]
{Patrick Weltevrede and Simon Johnston \\
Australia Telescope National Facility, CSIRO, P.O. Box 76, 
Epping, NSW 1710, Australia.
}
\date{Accepted 2008 April 23}
\pagerange{\pageref{firstpage}--\pageref{lastpage}}
\pubyear{2008}
\begin{document}
\maketitle
\label{firstpage}

\begin{abstract}
The observed fraction of pulsars with interpulses, their period
distribution and the observed pulse width versus pulse period
correlation is shown to be inconsistent with a model in which the
angle $\alpha$ between the magnetic axis and the rotation axis is
random. This conclusion appears to be unavoidable, even when
non-circular beams are considered. Allowing the magnetic axis to align
from a random distribution at birth with a timescale of
$\sim7\times10^7$ years can, however, explain those observations
well. The timescale derived is consistent with that obtained via
independent methods.  The probability that a pulsar beam intersects
the line of sight is a function of the angle $\alpha$ and therefore
beam evolution has important consequences for evolutionary models and
for estimations of the total number of neutron stars. The validity of
the standard formula for the spin-down rate, which is independent of
$\alpha$ appears to be questionable.
\end{abstract}


\section{Introduction}

Pulsars are formed from the supernova explosions which mark the death
of high mass stars. During the pulsar birth process, the pulsar's
rotation axis is aligned with its velocity vector
(\citealt{sp98,jhv+05}).  A further question arises as to the
orientation of the magnetic axis with respect to the rotation axis at
birth and the evolution of this angle with time. Over the years,
observational evidence seems to support the view that the magnetic and
rotation axes align over time (\citealt{cb86,lm88,tm98}), although
there have been claims for no evolution (\citealt{mck93,gh96}) and
even theoretical reasons for counter-alignment (\citealt{bgi88}).
Understanding this issue is important, not only for theoretical
considerations of how braking torque works in neutron stars, but also
for the evolution of the pulsar beam. This has important implications
for evolutionary models and for pulsar surveys generally, such as
those planned for the Square Kilometre Array (\citealt{kbl+04}).

Radio emission from pulsars arises from a few hundred kilometres above
the polar cap, the region bounded by the last open magnetic
field lines.  It can be shown simply that the radius of the polar cap,
$r$, is related to the pulsar spin period, $P$, via $r\propto
P^{-0.5}$.  In turn then, the opening angle of the beam, $\rho$, can
be expressed as
\begin{equation}
\label{EqH}
\rho = \sqrt{\frac{9\pi \,\,\, h_{\rm em}}{2\,\,\, P\,\,\, c}}
\end{equation}
(e.g \citealt{lk05}), 
where $h_{\rm em}$ is the emission height and $c$ the speed of light.

The region bounded by the last open magnetic  dipole field
lines, and therefore possibly also the pulsar beam, is only circular
if the magnetic axis is aligned with the rotation axis and for
non-aligned rotators the polar cap is compressed in the plain
containing the magnetic and the rotation axis
(e.g. \citealt{big90b}).  The beam shape becomes even more complex
when the distortion of the magnetic field lines close to the light
cylinder is considered (e.g. \citealt{ry95}). Also elongated beam
shapes in the latitudinal direction (e.g. \citealt{nv83}) as well as
an hourglass shape for the beam of the binary pulsar B1913+16
(\citealt{wt02}) have been proposed.

Observationally, one measures the pulse width ($\Delta\phi$) which is
a function of $\rho$ and the geometry of how the line of sight cuts
through the beam. Under the assumption that the beam is circular,
\cite{ggr84} derived
\begin{equation}
\label{EqW}
\sin^2\left(\frac{\Delta\phi}{4}\right) = \frac{\sin^2\left(\rho/2\right)-\sin^2\left(\beta/2\right)}{\sin\left(\alpha+\beta\right)\sin\alpha},
\end{equation}
where $\alpha$ is the angle between the magnetic axis and the rotation
axis and $\beta$ is the impact parameter, the angle between the line
of sight and the magnetic axis at its closest approach.
\cite{ran90} found that $\rho$ for core emission is correlated with
$P^{-1/2}$ by using pulsars with interpulses for which
$\alpha\simeq\degrees{90}$. Using this correlation in combination with
the longitude dependence of the position angle of the linear
polarization (\citealt{rc69a}) one can solve $\alpha$ and $\beta$ for
any pulsar with core emission. Several authors used this method to
derive a similar correlation between $\rho$ and $P$ for conal
emission (e.g. \citealt{ran93,kwj+94,gks93}). \cite{gou94} found
the same correlation by using the argument that the most narrow
profiles are expected for pulsars with $\alpha\simeq\degrees{90}$.
The implication of this
correlation in conjunction with Eq.~\ref{EqH} is that the emission
height and the fraction of the polar cap which is active is
independent of the pulsar period (e.g. \citealt{mr02a}).

\cite{tm98} analysed published values for $\alpha$ of several hundred
pulsars, showed they are inconsistent with a random $\alpha$
distribution and concluded that $\alpha$ decreases on a time-scale of
$\sim10^7$~yr. In contrast, \cite{mck93} concluded that the $\alpha$
distribution is random, provided that the beam was compressed in the
latitudinal direction.

In this paper, the issue of beam evolution is approached from a
different angle. We do not rely on polarization data or a $\rho$
correlation to determine $\alpha$ and $\beta$.
Rather, we will construct a beam model based on the observed fraction
of pulsars with interpulses and their period distribution and the
observed $P-\Delta\phi$ plane is used directly.
The interpulse population, the pulse width measurements and the model
are described in sections 2, 3 and 4 respectively. The results are
presented in section 5 and their implications are discussed section 6.

\section{\label{SctInterpulses}Interpulses}

The population of pulsars with interpulses is a powerful probe
of pulsar beam properties. In the case where an interpulse is seen,
with a separation from the main pulse of $\sim180\degr$ in
rotational phase, it implies that $\alpha$ must be $\sim90\degr$ and
therefore gives a good count of how many orthogonal rotators there are
in the population. We therefore searched the literature to obtain a
complete census of the interpulse population. For this search, we
excluded all millisecond pulsars and pulsars in globular clusters,
because the opening angle of the beams of milli-second pulsars are
believed to have a different $P$ dependence to that of normal pulsars
(e.g. \citealt{kxl+98}) and the values for the spin period derivative
($\dot{P}$) of globular cluster pulsars are contaminated by their
acceleration in the gravitational field of the cluster.  We started
with the catalogue of Taylor et al. (1993) which tabulated 14
interpulse pulsars out of a total of 527 objects. We then examined all
pulse profiles in the major surveys conducted since that time.

\begin{table}
\begin{center}
\begin{tabular}{l l r r}
Paper & Ref. & $N_\mathrm{Pulsars}$ & $N_\mathrm{IP}$\\
\hline
\cite{tml93} & (1) & 527 & 14\\
\cite{fcwa95} & & 14 & 0\\
\cite{cnst96} & & 19 & 0\\
\cite{dsb+98} & (2) & 84 & 1\\
\cite{ebvb01} & & 61 & 0\\
\cite{mlc+01} & & 99 & 0\\
\cite{mhl+02} & (3) & 119 & 3\\
\cite{kbm+03} & (4) & 200 & 2\\
\cite{hfs+04} & (5) & 178 & 2\\
\cite{jac05} & & 19 & 0\\
\cite{lxf+05} & & 8 & 0\\
\cite{bjd+06} &  & 14 & 0\\
\cite{lfl+06} & (6) & 131 & 2\\
\cite{mfl+06} & & 14 & 0\\
\cite{wj08b} & (7) & 0 & 3\\
\hline
Total & & 1487 & 27 \\
\end{tabular}
\end{center}
\caption{\label{TableSurveys}The used literature to find interpulses
is listed in the first column. The reference codes are the same as in
Table \ref{TableInterpulses}. $N_\mathrm{Pulsars}$ and
$N_\mathrm{IP}$ are the total number of newly discovered normal
pulsars in the discussed paper and the number of which have
interpulses.}
\end{table}

\begin{table}
\begin{center}
\begin{tabular}{l l c c c}
JName & BName & $P\,(s)$ & $\dot{P}$ & Ref.\\
\hline
J0534+2200  &    B0531+21   &     0.0331 & $4.2\times10^{-13}$   & (1) \\ 
J0826+2637  &    B0823+26   &     0.5307 & $1.7\times10^{-15}$   & (1) \\ 
J0828--3417  &    B0826--34   &     1.8489 & $1.0\times10^{-15}$ & (1) \\ 
J0834--4159  &               &     0.1211 & $4.4\times10^{-15}$  & (4) \\ 
J0905--5127  &               &     0.3463 & $2.5\times10^{-14}$  & (7) \\	 
J0908--4913  &    B0906--49   &     0.1068 & $1.5\times10^{-14}$ & (1) \\ 
J0953+0755  &    B0950+08   &     0.2531 & $2.3\times10^{-16}$   & (1) \\ 
J1057--5226  &    B1055--52   &     0.1971 & $5.8\times10^{-15}$ & (1) \\ 
J1126--6054  &    B1124--60   &     0.2027 & $2.8\times10^{-16}$ & (7) \\	 
J1302--6350  &    B1259--63   &     0.0478 & $2.3\times10^{-15}$ & (1) \\ 
J1549--4848  &               &     0.2883 & $1.4\times10^{-14}$  & (2) \\ 
J1637--4553  &    B1634--45   &     0.1188 & $3.2\times10^{-15}$ & (7) \\	 
J1705--1906  &    B1702--19   &     0.2990 & $4.1\times10^{-15}$ & (1) \\ 
J1713--3844  &               &     1.6001 & $1.8\times10^{-13}$  & (4) \\ 
J1722--3712  &    B1719--37   &     0.2362 & $1.1\times10^{-14}$ & (1) \\ 
J1739--2903  &    B1736--29   &     0.3229 & $7.9\times10^{-15}$ & (1) \\ 
J1806--1920  &               &     0.8798 & $1.7\times10^{-17}$  & (3) \\ 
J1808--1726  &               &     0.2410 & $1.2\times10^{-17}$  & (6) \\ 
J1825--0935  &    B1822--09   &     0.7690 & $5.2\times10^{-14}$ & (1) \\ 
J1828--1101  &               &     0.0721 & $1.5\times10^{-14}$  & (3) \\ 
J1843--0702  &               &     0.1916 & $2.1\times10^{-15}$  & (5) \\ 
J1849+0409  &               &     0.7612 & $2.2\times10^{-14}$   & (6) \\ 
J1851+0418  &    B1848+04   &     0.2847 & $1.1\times10^{-15}$   & (1) \\ 
J1852--0118  &               &     0.4515 & $1.8\times10^{-15}$  & (5) \\ 
J1913+0832  &               &     0.1344 & $4.6\times10^{-15}$   & (3) \\ 
J1932+1059  &    B1929+10   &     0.2265 & $1.2\times10^{-15}$   & (1) \\ 
J1946+1805  &    B1944+17   &     0.4406 & $2.4\times10^{-17}$   & (1) \\ 
\hline
\end{tabular}
\end{center}
\caption{\label{TableInterpulses}The 27 interpulses of normal
pulsars which can be found in the literature. The reference codes are
the same as in Table \ref{TableSurveys}.
}
\end{table}

Table \ref{TableSurveys} lists the relevant papers, the number of
pulsars reported and the interpulses detected. Table
\ref{TableInterpulses} lists the population of pulsars with
interpulses. The table includes 3 weak interpulses not previously
detected. Their profiles will be presented in a forthcoming
publication \citep{wj08b}. In total there are 27 pulsars with
interpulses from a total sample of 1487 pulsars, i.e. 1.8\% of the
population.  We note that the list should not be considered as
complete. Some objects in the list may not be orthogonal
rotators but are rather aligned rotators with wide beams (see
the debate for PSRs B0950+08 and B1929+10 in \citealt{ew01} for
example). On the other hand we may have missed some interpulse objects
because the flux in the interpulse is too weak to be detected.

\begin{figure}
\includegraphics[width=0.7\hsize,angle=270]{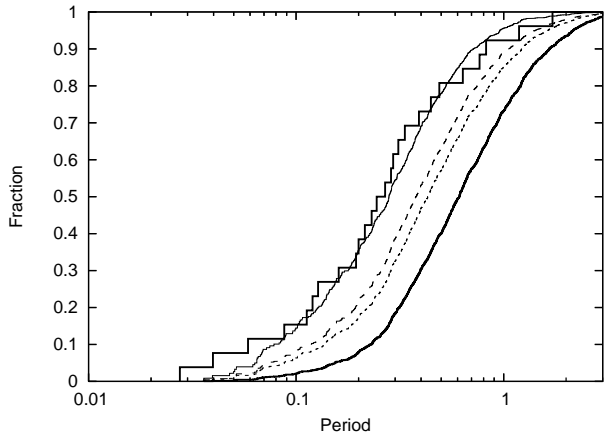}
\caption{\label{Fig_ip_period}The histogram (thick solid line) shows
the observed cumulative distribution of pulse periods of the
normal (i.e. non-millisecond) pulsars with interpulses. The lowest
(thick) solid curve shows the cumulative period distribution of all
normal pulsars. The dotted curve shows the predicted
distribution for the pulsars with interpulses in the case of a random
$\alpha$ distribution (both for circular and non-circular beams). The
dashed and solid curve are for a model including alignment of the
magnetic axis for non-circular and circular beams respectively.}
\end{figure}

If the $\alpha$ distribution of pulsars is random, then of the
order of 5\% of the pulsars can be expected to have an interpulse
(e.g. \citealt{gh96}). The measured percentage is lower, indicating a
potential problem for such a model. Not only is the percentage of
pulsars with interpulses an important clue for beam models, more
important still is their period distribution.  This distribution is
shown in Fig. \ref{Fig_ip_period} as a cumulative distribution (thick
solid histogram). The distribution indicates the fraction of pulsars
with interpulses $y$ which have a period less than $x$. Compared with
the period distribution of all normal (i.e. non-millisecond) pulsars
(lowest thick curve of Fig. \ref{Fig_ip_period}) there are relatively
many more short period pulsars with interpulses. In this paper we will
try to construct a geometrical model which can account for the
observed statistical properties of the population of pulsars with
interpulses.

\section{\label{SctPulseWidth}The observed pulse widths distribution}

We have embarked on a long-term timing campaign at the Parkes
telescope in Australia to monitor a large sample of  young pulsars
with high spin-down energy loss rates $\dot{E}$. This sample of
pulsars is compared with archival Parkes data, which is published in
various papers, in order to determine the differences between pulsars
with high and low values for $\dot{E}$ \citep{wj08b}. We refer to
that paper for the details of the observations and data reduction.

\begin{figure}
\includegraphics[width=0.7\hsize,angle=270]{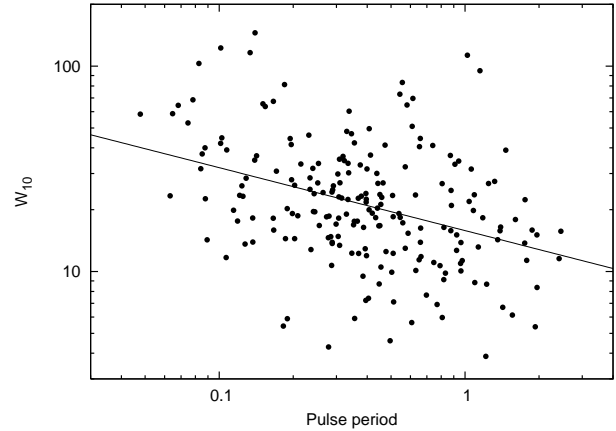}
\caption{\label{Fig_pulsewidths_p}The measured profile 10\% widths
versus $P$ at an observing frequency around $1390$ MHz
(\citealt{wj08b}). Pulsars with a $S/N < 30$ or those with substantial
scatter broadening were excluded leaving 205 measurements. The solid
line is a power law fit through the data, which has a slope of
$-0.31$, comparable with the fit obtained from the data of
\citet{gl98}.}
\end{figure}

The dataset is used to determine the correlation between the pulse
width $\Delta\phi$ and $P$. Fig. \ref{Fig_pulsewidths_p} suggests that
the slope in the $\Delta\phi-P$ plane is less steep than $-1/2$. The
slope measured by minimizing the $\chi^2$ in log-log space (assuming
equal weights) is $-0.31\pm0.05$. As pointed out in the introduction,
$\rho$ is expected, and found, to be proportional to $P^{-1/2}$. An
identical correlation can be expected in the $\Delta\phi-P$ plane,
provided that $\alpha$ is independent of $P$.  This measurement is
therefore of interest to us because it can be interpreted as evidence
for a $P$ dependence of $\alpha$.

One could argue that the scatter of the points around the correlation
in Fig. \ref{Fig_pulsewidths_p} is so large that a power law with a
slope of $-1/2$ would fit the data equally well. We therefore also
measured the slope using the measured 10\% widths at 1400 MHz by
\cite{gl98}, excluding the milli-second pulsars, as a consistency
check. Although there is some overlap in our samples, that of
\cite{gl98} contain many northern pulsars which are not present in our
sample. The slope measured in the \cite{gl98} data is also
flatter then expected ($-0.23\pm0.06$), which makes us more confident
that the observed trend is real.

\section{\label{SctModel}The model}

We attempt to explain three observational conclusions by creating a
model which is as simple as possible. If one understands the geometry
one should first of all be able to construct a model which can predict
the observed fraction of pulsars with an interpulse (1.8\% or lower,
see section 2). Secondly, the model should be able to reproduce the
observed period distribution of pulsars with interpulses. Thirdly, the
model should explain why the observed $\Delta\phi-P$ correlation is
flatter than the expected $P^{-1/2}$ correlation. The effects of a
possible alignment timescale of the magnetic axis and non-circular
beams will be explored.

The basic approach of this model is to synthesize a population of
pulsars in which the $P-\dot{P}$ distribution is identical to the
observed distribution. We then assign a random geometry to each pulsar
and the subpopulation of pulsars with interpulses is identified using
different model assumptions. The statistical properties of the
synthesized population of pulsars with interpulses is then compared
with the observed population in order to determine which model is able
to reproduce the data best.

\subsection{The period distribution}

The model should first of all reproduce the observed $P-\dot{P}$
distribution for the total observed population of pulsars (those with
and without interpulses), which is achieved by simply using the
observed $P-\dot{P}$ combinations. Not only does this ensure that the
resulting distributions are realistic, it also avoids the necessity to
make an full evolutionary model which would require additional
assumptions about pulsar birth properties and their evolution. 
Note that we are therefore not making any assumptions about the parent
distribution that gives rise to the observed distribution of
$P-\dot{P}$ combinations. In the next subsection we will describe
geometrical factors which could make the period distribution of the
pulsars with interpulses different to the total population of observed
pulsars.

The catalogue of pulsars maintained by the
ATNF\footnote{http://www.atnf.csiro.au/research/pulsar/psrcat}
\citep{mhth05} was used to obtain the observed $P-\dot{P}$ values.
Because the interpulses of milli-second pulsars are not
considered in this paper, all pulsars with an estimated surface
magnetic field strength below $10^{11}$ Gauss or a period faster than
the Crab were excluded. PSR J1808--1726, with a magnetic field
strength of $5.4\times10^{10}$ Gauss, is excluded from the interpulse
statistics for consistency. The globular cluster pulsars have
unreliable $\dot{P}$ values and are therefore also not considered.
Finally, all the pulsars with $P>4$ seconds were removed to exclude
the catalogued soft gamma-ray repeaters and the anomalous X-ray
pulsars.

\subsection{The beam model}

The first step is to draw random $P-\dot{P}$ pairs from the total
population of observed pulsars as described in Sect 4.1. Secondly,
random values are generated for $\alpha$ and $\beta$. The magnetic
pole and the line of sight are taken to be a random points on a
sphere. This implies that the distribution of $\alpha$ and
$\zeta=\alpha+\beta$ are sinusoidal.  The magnetic axis is allowed to
align on a timescale $\tau_\mathrm{align}$ (\citealt{jon76})
\begin{equation}
\alpha(t) = \alpha_0\exp\left(-t/\tau_\mathrm{align}\right),
\end{equation}
where $\alpha_0$ is the random $\alpha$ value at birth and $t$ is the
age of the pulsar, which is taken to be the characteristic age. The
final step is to determine if one or both beams of the pulsar
intersect the line of sight given by the randomly generated geometry,
which depends on the assumed beam shape. As pointed out in the
introduction, different shapes of the polar cap have been
proposed. Here we consider only the shape of a polar cap which is
bounded by the last open magnetic dipole field lines in order to get a
feeling for the effect of non-circular beam shapes. Such a beam is
compressed in the plain containing the magnetic and the rotation axis
and is roughly elliptic. The axial ratio $\cal{E}(\alpha)$ of the
ellipse is given by (\citealt{mck93})
\begin{equation}
\label{Eqepsilon}
\cal{E}(\alpha) = \cos\delta\sqrt{\cos\left(\alpha-\delta\right)},
\end{equation}
where $\delta(\alpha)$ follows from numerically solving 
\begin{equation}
2\tan\left(\delta\right) = \tan\left(\alpha-\delta\right).
\end{equation}
The axial ratio $\cal{E}(\alpha)$ varies from 1 (for an aligned beam)
to $\sim0.62$ (for an orthogonal beam). Circular beams can be assumed
in the model by forcing $\cal{E}=\mathrm{1}$ for all $\alpha$.

Apart from the beam shape, a beam size has also to be assumed. For the
half opening angle of the beam we use the relation measured by
\cite{gou94}
\begin{equation}
\label{Eqrho}
\rho=\degrees{5.4}P^{-1/2},
\end{equation}
which is consistent with the relation reported by various other
authors (e.g. \citealt{ran90,ran93,kwj+94,gks93}). For non-circular
beams, $\rho$ of Eq. \ref{Eqrho} is taken to be the half opening angle
in the longitudinal direction. The conditions for the two pulsar beams
to intersect the line of sight are
\begin{equation}
\label{condition1}
\left|\beta\right|\le \cal{E}(\alpha)\,\rho
\end{equation}
and
\begin{equation}
\label{condition2}
\left|\degrees{180}-2\alpha-\beta\right|\le \cal{E}(\alpha)\,\rho.
\end{equation}
Random values of $\alpha$ and $\beta$ are generated until at least one
of these conditions are met. The width of the pulse profile can then
be calculated with Eq. \ref{EqW} for circular beams.

As noted in section 2, the list of interpulses is not complete
for two reasons. The first reason is that some interpulses
might have been missed because the flux in the interpulse is too weak
to be detected. Extremely weak interpulses could indicate that the
pulsar beam just grazes the line of sight. Because the observed $\rho$
(Eq. \ref{Eqrho}) has been derived considering the 10\% pulse widths,
beams which just graze the line of sight are not counted as detections
in the model (Eqs. \ref{condition1} and \ref{condition2}). This means
that extremely weak interpulses are not accounted for in the model, as
well as possibly in the observations. However, some of the observed
interpulses will be wide beams of aligned rotators. Therefore the
observed interpulse fraction should be considered as an upper
limit.

\section{\label{SctResults}Results}

Let us start with considering the simplest model, i.e. a random
$\alpha$ distribution without alignment ($\tau_\mathrm{align}=\infty$)
and circular beams ($\cal{E}(\alpha)=\mathrm{1}$). In this
scenario our model predicts that 4.4\% of the pulsars should have an
interpulse, a factor of at least two higher than the observed
fraction. This model also clearly fails to explain the observed period
distribution of the pulsars with interpulses as the dotted curve of
Fig. \ref{Fig_ip_period} lies below the observed distribution. This
means that there are more short period pulsars with interpulses then
predicted. Short period pulsars are expected to have a larger
probability to show an interpulse because their beams are wider
(Eq. \ref{Eqrho}), which is the reason why the predicted distribution
of pulsars with interpulses (dotted line) lies above the distribution
of all pulsars (lowest thick curve). Nevertheless this effect is not
enough to explain the observed period distribution of the pulsars with
interpulses. The predicted pulse width for this model is correlated
with $P^{-0.51}$, very close to the what is expected from
Eq. \ref{Eqrho}. This model therefore fails to explain the three
observational properties outlined in section 4.

In order to make the model more accurately reflect the data it is
necessary to produce fewer pulsars with interpulses, which can be done
by relaxing the condition that forces the beams to be
circular. Because the minor axis of non-circular beams is in the
plane containing the magnetic and rotation axis, it is less likely
that the beams of both magnetic poles intersect the line of
sight. This in contrast to elongated beams, which will increase the
fraction of pulsars with interpulses and are therefore not considered
further here.  Indeed, when applying Eq. \ref{Eqepsilon}, the
predicted fraction of interpulses is decreased to 2.3\%. This is still
a bit too high, especially as the observed fraction is an upper
limit. Because the ellipticity of the beam is independent of $P$ it
follows that the predicted pulse period distribution of pulsars with
interpulses and the pulse width distribution are identical to the
model with circular beams. Therefore the model with non-circular beams
also fails to explain the observations.

We next consider the effect of alignment of the pulsar beam. By
setting $\tau_\mathrm{align}=7\times10^7$ years (assuming circular
beams) the predicted fraction of pulsars with interpulses drops to the
observed value. Also the predicted period distribution of the pulsars
with interpulses (thin solid line Fig. \ref{Fig_ip_period}) now agrees
well with the observations, indicating that the model successfully
predicts the observed under-abundance of long period pulsars with
interpulses. This can be understood because long period pulsars tend
to have larger characteristic ages, which means that they tend to be
more aligned, which reduces the probability that the beams of both
magnetic poles intersect the line of sight. Finally, because the
$\alpha$ distribution depends on $P$ in this case, the predicted
correlation between $\Delta\phi$ and $P$ will deviate from $-1/2$. The
slope of the correlation becomes flatter (the slope is $-0.40$), as is
observed.

Finally, one can consider a model which both includes the ellipticity
of the beam and alignment. In that case the timescale of alignment is
much longer ($\tau_\mathrm{align}=2\times10^9$), because both
effects reduce the predicted number of interpulses. As a
consequence, the predicted period distribution of the pulsars with
interpulses (dashed line Fig. \ref{Fig_ip_period}) does not deviate
much from the model without alignment and therefore does not fit the
data as well as the model with circular beams.

If pulsar beams align over time, it would imply that the total
population of pulsars has an $\alpha$ distribution which is skewed to
low values compared with a random distribution. This affects the
average beaming fraction because it is a function of $\alpha$ 
(e.g. Eq. 7 of \citealt{tm98}). The beaming fraction is the fraction
of the celestial sphere swept out by the beams of a pulsar, hence it
is the probability that a pulsar is observable for a random line of
sight. Aligned beams are less likely to intersect the line of sight
and are therefore less likely to be observed. The average beaming
fraction for the observed population of pulsars, assuming circular
beams, is 8.4\% and 17\% for the model with and without $\alpha$
evolution respectively.

For a random $\alpha$ distribution the fraction of pulsars with beams
which continuously intersect the line of sight is tiny (0.01\%),
because of the $\alpha$ dependence of the beaming fraction in
combination with the effect that aligned pulsars are less likely to be
formed in the first place. This means that basically no pulsars with
extremely wide profiles are expected in the known population of
pulsars.  This fraction appears too low, as we have already argued
that some of the interpulses in Table \ref{TableInterpulses} originate
from a single magnetic pole. A more realistic fraction is predicted by
the model assuming circular beams including $\alpha$ evolution
(1.6\%), while $\tau_\mathrm{align}$ is too long to make a
difference in the case of non-circular beams. This can be seen as
additional evidence for circular beams and a significant effect of
$\alpha$ evolution.

\section{\label{SctDiscussion}Discussion}

We have shown that the observed fraction of pulsars with interpulses,
their period distribution, the fraction of pulsars with extremely
wide profiles and the observed pulse width versus pulse period
correlation (as measured in our data and that of \citealt{gl98}) is
inconsistent with a random $\alpha$ distribution, even when
non-circular beams are considered. The main problem is that too many
pulsars with interpulses are predicted, especially those with long
periods. A number of explanations are possible.

First, one could assume that long period pulsars are less
likely to have two active poles. Not only is this assumption ad-hoc,
it also does not explain the observed correlation between the pulse
width and the pulse period. Our model is kept simple and intuitive by
assuming that both magnetic poles always emit radio emission, the
beams are directed in exactly opposite directions and have equal
widths and luminosities. We therefore aim to find a geometrical
solution for the problem.

Secondly, one could assume a different powerlaw relation between
$\rho$ and $P$. In order to correctly predict the number of pulsars
with interpulses and their period distribution the observed powerlaw
relation (Eq. \ref{Eqrho}) should be steeper and the constant
smaller. This could physically mean that the fraction of the polar
cap which is active or the emission height is $P$ dependent. Not only
is such a model incompatible with the observations from which
Eq. \ref{Eqrho} is derived, it also results in a highly unrealistic
pulse width distribution (the predicted pulse widths are too narrow
and the correlation with $P$ is too steep).

Thirdly, the pulsar beam may not be circular, which is expected if the
polar cap is bounded by the last open magnetic field lines.
Although elliptical beams which are compressed in the plain containing
the magnetic and the rotation axis could help to reduce the number of
predicted interpulses, it is not enough. Moreover, the ellipticity
is not expected to be $P$ dependent. This implies that this model,
like that for circular beams, does not correctly predict the period
distribution of the pulsars interpulses and the pulse width period
correlation. The fraction of pulsars with extemely wide profiles is
also too low.

The only geometrical effect left, as far as we can see, to explain the
observations is to relax the assumption that $\alpha$ is random. In
order to explain the observed fraction of pulsars with interpulses the
$\alpha$ distribution should be skewed to low values. The distribution
for short period pulsars should be skewed less than that of long
period pulsars in order to explain the observed pulse period
distribution of the pulsars with interpulses. The skewness therefore
appears to be caused by an {\em evolution in time} of $\alpha$ rather
than by a non-random birth distribution which is fixed in time.

In the previous section it has been shown that the model assuming
circular beams can explain the observations well when the magnetic
axis is allowed to align from a random distribution at birth with a
timescale of $\sim7\times10^7$ years.  The fact that the fraction of
pulsars with interpulses, their period dependence, the fraction
of pulsars with extremely wide profiles and the pulse width
distribution can be explained by adding just one extra model parameter
is encouraging. Moreover, by allowing $\alpha$ to evolve it is not
necessary to abandon the observed $\rho-P$ relation (Eq. \ref{Eqrho})
or to resort in ad-hoc assumptions about magnetic poles which are not
always active.

The timescale is comparible with the timescale derived from the
observed $\alpha$ distribution by \cite{tm98}.  Measuring $\alpha$ is
far from trivial and for most pulsars it is, at best, only poorly
constrained. It should be stressed that although we discuss $\alpha$
evolution, we do not rely on (indirect) measurements of
$\alpha$. Therefore our results are an important addition to the
existing evidence for alignment.

The pulse period distribution of the pulsars with interpulses can be
used to distinguish between beam shapes. The data show that the
derived distribution assuming circular beams fits the data better than
that assuming elliptical beams. This suggests that the beams are
more circular than expected from a simple model which has the polar
cap bounded by the last open dipole magnetic field lines.  For
example, more complicated models which take into account the
distortion of the magnetic field lines close to the light cylinder
predict roughly circular beams (e.g. \citealt{ry95}).

Our estimation of $\tau_\mathrm{align}$ should be considered to be
very rough because of a number of reasons. First of all, the timescale
is based on the observed characteristic age distribution of
pulsars. The characteristic age should be considered to be only a
rough estimator of the true age, because the details of pulsar
spin-down are not well understood. Secondly, the $\alpha$ evolution
could have a different form than the assumed exponential decay.  For
instance one could imagine that the decay rate is a function of
$\alpha$. Thirdly, $\tau_\mathrm{align}$ also depends on the shape of
the radio beam. If the beam is non-circular then
$\tau_\mathrm{align}$ should be larger in order to explain the
observations. Finally, the timescale is based on the observed fraction
of interpulses. It is assumed that all the observed interpulses are
caused by emission from the opposite pole, which is unlikely to be the
case. Therefore the observed fraction of pulsars with interpulses
should be seen as an upper limit, resulting in an overestimation of
$\tau_\mathrm{align}$.

Although our estimated value of $\tau_\mathrm{align}$ is very
rough, the observations can only be explained if the effect of
alignment is important and therefore happens on a timescale comparable
to the age of a typical pulsar. Beam evolution has some important
consequences for population studies, evolutionary models and pulsar
searches, for two reasons.

The first is that the beaming fraction depends on
$\tau_\mathrm{align}$ such that aligned beams are less likely to
intersect the line of sight. Therefore a consequence of alignment of
pulsar beams over time is that older pulsars become harder to find,
hence the inferred total population of old pulsars should be larger.
The average beaming fraction for the observed population of pulsars,
assuming circular beams, is 8.4\% and 17\% for the model with and
without $\alpha$ evolution respectively. Therefore the inferred total
population of pulsars is about twice as high when $\alpha$ evolution
is considered.

Secondly, one can ask the more fundamental question of what alignment
implies for the spin-down of pulsars. The evolution of $\alpha$ must
be the result of the braking torque, hence it is related to the
spin-down of pulsars. If the torque changes $\alpha$, it seems
reasonable that the torque also depends on $\alpha$. It is therefore
not unlikely that the pulsar spin-down itself has an $\alpha$
dependence. The spin-down of a rotating dipole field in a vacuum
depends on $\alpha$ ($\dot{E}\propto\sin^2\alpha$,
e.g. \citealt{jac62}), but how different this is in the presence of a
pulsar magnetosphere is unclear. An $\alpha$ dependent spin-down
implies that the magnetic field strength, the characteristic age and
spin-down energy loss rate of a pulsar also depends on $\alpha$, which
has important consequences for evolutionary models and population
studies.

We have shown that the observed population of pulsars with interpulses
places strong restrictions on beam models. Future instrumentation,
such as the Square Kilometre Array, will greatly enlarge the observed
population of pulsars, enabling the possibility to distinguish between
models with different beam shapes and different functional forms for
the $\alpha$ evolution. This will contribute to a better understanding
of how breaking torques in neutron stars work.

\section*{Acknowledgments}
The Australia Telescope is funded by the Commonwealth of Australia for
operation as a National Facility managed by the CSIRO. We would like
to thank the referee for his/her constructive comments.

\bibliographystyle{mn2e}
\bibliography{journals_apj,modrefs,psrrefs,crossrefs} 

\begin{thebibliography}{}

\bibitem[\protect\citeauthoryear{Beskin, Gurevich \& Istomin}{Beskin
  et~al.}{1988}]{bgi88}
Beskin V.~S.,  Gurevich A.~V.,    Istomin Y.~N.,  1988, Ap\&SS, 146, 205

\bibitem[\protect\citeauthoryear{Biggs}{Biggs}{1990}]{big90b}
Biggs J.~D.,  1990, MNRAS, 245, 514

\bibitem[\protect\citeauthoryear{{Burgay}, {Joshi}, {D'Amico}, {Possenti},
  {Lyne}, {Manchester}, {McLaughlin}, {Kramer}, {Camilo} \& {Freire}}{{Burgay}
  et~al.}{2006}]{bjd+06}
{Burgay} M.,  {Joshi} B.~C.,  {D'Amico} N.,  {Possenti} A.,  {Lyne} A.~G.,
  {Manchester} R.~N.,  {McLaughlin} M.~A.,  {Kramer} M.,  {Camilo} F.,
  {Freire} P.~C.~C.,  2006, MNRAS, 368, 283

\bibitem[\protect\citeauthoryear{Camilo, Nice, Shrauner \& Taylor}{Camilo
  et~al.}{1996}]{cnst96}
Camilo F.,  Nice D.~J.,  Shrauner J.~A.,    Taylor J.~H.,  1996, ApJ, 469, 819

\bibitem[\protect\citeauthoryear{{Candy} \& {Blair}}{{Candy} \&
  {Blair}}{1986}]{cb86}
{Candy} B.~N.,  {Blair} D.~G.,  1986, ApJ, 307, 535

\bibitem[\protect\citeauthoryear{D'Amico, Stappers, Bailes, Martin, Bell, Lyne
  \& Manchester}{D'Amico et~al.}{1998}]{dsb+98}
D'Amico N.,  Stappers B.~W.,  Bailes M.,  Martin C.~E.,  Bell J.~F.,  Lyne
  A.~G.,    Manchester R.~N.,  1998, MNRAS, 297, 28

\bibitem[\protect\citeauthoryear{{Edwards}, {Bailes}, {van Straten} \&
  {Britton}}{{Edwards} et~al.}{2001}]{ebvb01}
{Edwards} R.~T.,  {Bailes} M.,  {van Straten} W.,    {Britton} M.~C.,  2001,
  MNRAS, 326, 358

\bibitem[\protect\citeauthoryear{{Everett} \& {Weisberg}}{{Everett} \&
  {Weisberg}}{2001}]{ew01}
{Everett} J.~E.,  {Weisberg} J.~M.,  2001, ApJ, 553, 341

\bibitem[\protect\citeauthoryear{Foster, Cadwell, Wolszczan \& Anderson}{Foster
  et~al.}{1995}]{fcwa95}
Foster R.~S.,  Cadwell B.~J.,  Wolszczan A.,    Anderson S.~B.,  1995, ApJ,
  454, 826

\bibitem[\protect\citeauthoryear{Gil, Gronkowski \& Rudnicki}{Gil
  et~al.}{1984}]{ggr84}
Gil J.~A.,  Gronkowski P.,    Rudnicki W.,  1984, A\&A, 132, 312

\bibitem[\protect\citeauthoryear{Gil \& Han}{Gil \& Han}{1996}]{gh96}
Gil J.~A.,  Han J.~L.,  1996, ApJ, 458, 265

\bibitem[\protect\citeauthoryear{Gil, Kijak \& Seiradakis}{Gil
  et~al.}{1993}]{gks93}
Gil J.~A.,  Kijak J.,    Seiradakis J.~H.,  1993, A\&A, 272, 268

\bibitem[\protect\citeauthoryear{Gould}{Gould}{1994}]{gou94}
Gould D.~M.,  1994, PhD thesis, University of Manchester

\bibitem[\protect\citeauthoryear{Gould \& Lyne}{Gould \& Lyne}{1998}]{gl98}
Gould D.~M.,  Lyne A.~G.,  1998, MNRAS, 301, 235

\bibitem[\protect\citeauthoryear{{Hobbs}, {Faulkner}, {Stairs}, {Camilo},
  {Manchester}, {Lyne}, {Kramer}, {D'Amico}, {Kaspi}, {Possenti}, {McLaughlin},
  {Lorimer}, {Burgay}, {Joshi} \& {Crawford}}{{Hobbs} et~al.}{2004}]{hfs+04}
{Hobbs} G.,  {Faulkner} A.,  {Stairs} I.~H.,  {Camilo} F.,  {Manchester} R.~N.,
   {Lyne} A.~G.,  {Kramer} M.,  {D'Amico} N.,  {Kaspi} V.~M.,  {Possenti} A.,
  {McLaughlin} M.~A.,  {Lorimer} D.~R.,  {Burgay} M.,  {Joshi} B.~C.,
  {Crawford} F.,  2004, MNRAS, 352, 1439

\bibitem[\protect\citeauthoryear{Jackson}{Jackson}{1962}]{jac62}
Jackson J.~D.,  1962, Classical Electrodynamics.
Wiley

\bibitem[\protect\citeauthoryear{{Jacoby}}{{Jacoby}}{2005}]{jac05}
{Jacoby} B.~A.,  2005, PhD thesis, California Institute of Technology, United
  States -- California

\bibitem[\protect\citeauthoryear{{Johnston}, {Hobbs}, {Vigeland}, {Kramer},
  {Weisberg} \& {Lyne}}{{Johnston} et~al.}{2005}]{jhv+05}
{Johnston} S.,  {Hobbs} G.,  {Vigeland} S.,  {Kramer} M.,  {Weisberg} J.~M.,
  {Lyne} A.~G.,  2005, MNRAS, 364, 1397

\bibitem[\protect\citeauthoryear{{Jones}}{{Jones}}{1976}]{jon76}
{Jones} P.~B.,  1976, Astrophys. Space Sci., 45, 369

\bibitem[\protect\citeauthoryear{Kramer, Backer, Lazio, Stappers \&
  Johnston}{Kramer et~al.}{2004}]{kbl+04}
Kramer M.,  Backer D.~C.,  Lazio T. J.~W.,  Stappers B.~W.,    Johnston S.,
  2004, New Astron. Rev., 48, 993

\bibitem[\protect\citeauthoryear{{Kramer}, {Bell}, {Manchester}, {Lyne},
  {Camilo}, {Stairs}, {D'Amico}, {Kaspi}, {Hobbs}, {Morris}, {Crawford},
  {Possenti}, {Joshi}, {McLaughlin}, {Lorimer} \& {Faulkner}}{{Kramer}
  et~al.}{2003}]{kbm+03}
{Kramer} M.,  {Bell} J.~F.,  {Manchester} R.~N.,  {Lyne} A.~G.,  {Camilo} F.,
  {Stairs} I.~H.,  {D'Amico} N.,  {Kaspi} V.~M.,  {Hobbs} G.,  {Morris} D.~J.,
  {Crawford} F.,  {Possenti} A.,  {Joshi} B.~C.,  {McLaughlin} M.~A.,
  {Lorimer} D.~R.,    {Faulkner} A.~J.,  2003, MNRAS, 342, 1299

\bibitem[\protect\citeauthoryear{Kramer, Wielebinski, Jessner, Gil \&
  Seiradakis}{Kramer et~al.}{1994}]{kwj+94}
Kramer M.,  Wielebinski R.,  Jessner A.,  Gil J.~A.,    Seiradakis J.~H.,
  1994, A\&AS, 107, 515

\bibitem[\protect\citeauthoryear{Kramer, Xilouris, Lorimer, Doroshenko,
  Jessner, Wielebinski, Wolszczan \& Camilo}{Kramer et~al.}{1998}]{kxl+98}
Kramer M.,  Xilouris K.~M.,  Lorimer D.~R.,  Doroshenko O.,  Jessner A.,
  Wielebinski R.,  Wolszczan A.,    Camilo F.,  1998, ApJ, 501, 270

\bibitem[\protect\citeauthoryear{{Lorimer}, {Faulkner}, {Lyne}, {Manchester},
  {Kramer}, {McLaughlin}, {Hobbs}, {Possenti}, {Stairs}, {Camilo}, {Burgay},
  {D'Amico}, {Corongiu} \& {Crawford}}{{Lorimer} et~al.}{2006}]{lfl+06}
{Lorimer} D.~R.,  {Faulkner} A.~J.,  {Lyne} A.~G.,  {Manchester} R.~N.,
  {Kramer} M.,  {McLaughlin} M.~A.,  {Hobbs} G.,  {Possenti} A.,  {Stairs}
  I.~H.,  {Camilo} F.,  {Burgay} M.,  {D'Amico} N.,  {Corongiu} A.,
  {Crawford} F.,  2006, MNRAS, 372, 777

\bibitem[\protect\citeauthoryear{{Lorimer} \& {Kramer}}{{Lorimer} \&
  {Kramer}}{2005}]{lk05}
{Lorimer} D.~R.,  {Kramer} M.,  2005, {Handbook of Pulsar Astronomy}.
Cambridge University Press

\bibitem[\protect\citeauthoryear{{Lorimer}, {Xilouris}, {Fruchter}, {Stairs},
  {Camilo}, {Vazquez}, {Eder}, {McLaughlin}, {Roberts}, {Hessels} \&
  {Ransom}}{{Lorimer} et~al.}{2005}]{lxf+05}
{Lorimer} D.~R.,  {Xilouris} K.~M.,  {Fruchter} A.~S.,  {Stairs} I.~H.,
  {Camilo} F.,  {Vazquez} A.~M.,  {Eder} J.~A.,  {McLaughlin} M.~A.,  {Roberts}
  M.~S.~E.,  {Hessels} J.~W.~T.,    {Ransom} S.~M.,  2005, MNRAS, 359, 1524

\bibitem[\protect\citeauthoryear{Lyne \& Manchester}{Lyne \&
  Manchester}{1988}]{lm88}
Lyne A.~G.,  Manchester R.~N.,  1988, MNRAS, 234, 477

\bibitem[\protect\citeauthoryear{{Manchester}, {Fan}, {Lyne}, {Kaspi} \&
  {Crawford}}{{Manchester} et~al.}{2006}]{mfl+06}
{Manchester} R.~N.,  {Fan} G.,  {Lyne} A.~G.,  {Kaspi} V.~M.,    {Crawford} F.,
   2006, ApJ, 649, 235

\bibitem[\protect\citeauthoryear{Manchester, Hobbs, Teoh \& Hobbs}{Manchester
  et~al.}{2005}]{mhth05}
Manchester R.~N.,  Hobbs G.~B.,  Teoh A.,    Hobbs M.,  2005, AJ, 129, 1993

\bibitem[\protect\citeauthoryear{Manchester, Lyne, Camilo, Bell, Kaspi,
  D'Amico, McKay, Crawford, Stairs, Possenti, Morris \& Sheppard}{Manchester
  et~al.}{2001}]{mlc+01}
Manchester R.~N.,  Lyne A.~G.,  Camilo F.,  Bell J.~F.,  Kaspi V.~M.,  D'Amico
  N.,  McKay N. P.~F.,  Crawford F.,  Stairs I.~H.,  Possenti A.,  Morris
  D.~J.,    Sheppard D.~C.,  2001, MNRAS, 328, 17

\bibitem[\protect\citeauthoryear{{McKinnon}}{{McKinnon}}{1993}]{mck93}
{McKinnon} M.~M.,  1993, ApJ, 413, 317

\bibitem[\protect\citeauthoryear{{Mitra} \& {Rankin}}{{Mitra} \&
  {Rankin}}{2002}]{mr02a}
{Mitra} D.,  {Rankin} J.~M.,  2002, ApJ, 577, 322

\bibitem[\protect\citeauthoryear{{Morris}, {Hobbs}, {Lyne}, {Stairs}, {Camilo},
  {Manchester}, {Possenti}, {Bell}, {Kaspi}, {Amico}, {McKay}, {Crawford} \&
  {Kramer}}{{Morris} et~al.}{2002}]{mhl+02}
{Morris} D.~J.,  {Hobbs} G.,  {Lyne} A.~G.,  {Stairs} I.~H.,  {Camilo} F.,
  {Manchester} R.~N.,  {Possenti} A.,  {Bell} J.~F.,  {Kaspi} V.~M.,  {Amico}
  N.~D.,  {McKay} N.~P.~F.,  {Crawford} F.,    {Kramer} M.,  2002, MNRAS, 335,
  275

\bibitem[\protect\citeauthoryear{Narayan \& Vivekanand}{Narayan \&
  Vivekanand}{1983}]{nv83}
Narayan R.,  Vivekanand M.,  1983, A\&A, 122, 45

\bibitem[\protect\citeauthoryear{Radhakrishnan \& Cooke}{Radhakrishnan \&
  Cooke}{1969}]{rc69a}
Radhakrishnan V.,  Cooke D.~J.,  1969, Astrophys. Lett., 3, 225

\bibitem[\protect\citeauthoryear{Rankin}{Rankin}{1990}]{ran90}
Rankin J.~M.,  1990, ApJ, 352, 247

\bibitem[\protect\citeauthoryear{Rankin}{Rankin}{1993}]{ran93}
Rankin J.~M.,  1993, ApJ, 405, 285

\bibitem[\protect\citeauthoryear{Romani \& Yadigaroglu}{Romani \&
  Yadigaroglu}{1995}]{ry95}
Romani R.~W.,  Yadigaroglu I.-A.,  1995, ApJ, 438, 314

\bibitem[\protect\citeauthoryear{Spruit \& Phinney}{Spruit \&
  Phinney}{1998}]{sp98}
Spruit H.,  Phinney E.~S.,  1998, Nature, 393, 139

\bibitem[\protect\citeauthoryear{Tauris \& Manchester}{Tauris \&
  Manchester}{1998}]{tm98}
Tauris T.~M.,  Manchester R.~N.,  1998, MNRAS, 298, 625

\bibitem[\protect\citeauthoryear{Taylor, Manchester \& Lyne}{Taylor
  et~al.}{1993}]{tml93}
Taylor J.~H.,  Manchester R.~N.,    Lyne A.~G.,  1993, ApJS, 88, 529

\bibitem[\protect\citeauthoryear{{Weisberg} \& {Taylor}}{{Weisberg} \&
  {Taylor}}{2002}]{wt02}
{Weisberg} J.~M.,  {Taylor} J.~H.,  2002, ApJ, 576, 942

\bibitem[\protect\citeauthoryear{{Weltevrede} \& {Johnston}}{{Weltevrede} \&
  {Johnston}}{2008}]{wj08b}
{Weltevrede} P.,  {Johnston} S.,  2008, MNRAS, in prep.

\end{thebibliography}
\label{lastpage} 

\clearpage
\end{document}